\documentclass{article}
\usepackage{authblk} 
\usepackage{hyperref} 
\usepackage{amsmath}
\usepackage{graphicx}

\usepackage{comment}

\def\Sdk#1#2{{}_{#1}S_{#2}}
\def\Xdk#1#2{{}_{#1}X_{#2}}
\def\Pdk#1#2{{}_{#1}P_{#2}}
\def\pdk#1#2{{}_{\scriptstyle{#1}}{{\raisebox{0.4ex}{$p$}}}_{\scriptstyle{_{\!}#2}}}
\def\ppdk#1#2{{}_{\scriptstyle{#1}}{{\raisebox{0.4ex}{$\tilde{p}$}}}_{\scriptstyle{_{\!}#2}}}
\def\sdk#1#2{{}_{\scriptstyle{#1}_{\!}}{{\raisebox{0.25ex}{$s$}}}_{\scriptstyle{_{\!}#2}}}
\def\sab#1{{\raisebox{0.15ex}{$s$}}_{\scriptscriptstyle{\!#1}}}
\def\delm{\Delta^{k_1,k_2}_\text{max}(D_\text{th})}
\def\delmb{\Delta^{k_2,k_1}_\text{max}(D_\text{th})}
\def\delme#1#2#3{\Delta^{#1,#2}_\text{max}(#3)}
\def\deld{\Delta^{k_1,k_2}_\text{diff}(D_\text{th})}
\def\delde#1#2#3{\Delta^{#1,#2}_\text{diff}(#3)}

\def\beqa{\begin{eqnarray}}
\def\eeqa{\end{eqnarray}}
\def\a={&=&}

\newcommand{\pa}[1]{{\left( {#1} \right)}}

\title{Patterns and Dynamics of Netflix TV Show Popularity}

\author[1]{Nahyeon Lee}
\author[2]{Jongsoo Lim}
\author[3,4]{Hyeong-Chai Jeong\thanks{\href{mailto:hcj@sejong.ac.kr}{hcj@sejong.ac.kr}}}

\affil[1]{Department of Physics, Sejong University, Seoul 05006, Republic of Korea}
\affil[2]{Department of Media and Communication, Sejong University, Seoul 05006, Republic of Korea}
\affil[3]{Department of Physics and Astronomy, Sejong University, Seoul 05006, Republic of Korea}
\affil[4]{School of Computational Sciences, Korea Institute for Advanced Study, Seoul 02455, Republic of Korea}

\date{} 

\begin{document}

\maketitle

\begin{abstract}\label{sec1}
The rise of platforms like Netflix has expanded the possibility for audiences worldwide to watch the same content simultaneously, motivating research on cross-country media consumption.
We investigate the global dynamics of media consumption by analyzing daily top-ranked Netflix TV shows across 71 countries over a span of 822 days.
Using an information-theoretic framework, we measure diversity, similarity, and directional relationships in consumption trends using Shannon entropy, mutual information, and Kullback-Leibler (KL) divergence.
According to Shannon entropy analysis, North America and Europe have highly dynamic viewing preferences, whereas East and Southeast Asia (ESA) display more persistent trends, with the same shows often dominating for long periods.
Mutual information identifies clear regional clusters of synchronized consumption, with particularly strong alignment among ESA countries. 
To analyze temporal patterns, we introduce a KL-based asymmetry measure that captures directional patterns between countries, applicable to both inter- and intra-regional pairs.
This analysis reveals distinct pathways of content spread.
We find inter-regional patterns from ESA and South America toward North America and Europe, and intra-regional signals from Korea and Thailand to other ESA countries.
We also observe that ESA trends reaching other regions often originate from Singapore.
These findings offer insight into the temporal structure of global content spread and highlight the coexistence of global synchronization and regional independence in streaming media preferences.
\end{abstract}

\section{Introduction}
The global growth of streaming services such as Netflix, Amazon Prime and Disney+ has made digital media consumption an important force in international cultural exchange.
Based on this development, many studies have focused on content diversity, language similarity, and audience segmentation~\cite{afilipoaie2021netflix, wayne2024loved, park2025social}. 
Some have grouped countries based on cultural or linguistic similarity, although their group compositions vary across studies~\cite{lee2025global,jang2023global}.
Other works have explored how streaming services influence viewer behavior and content engagement~\cite{lotz2021between,lee2024strategies}, and how recommendation systems adapt to the genre and topic preferences of specific countries~\cite{cornelio2020mexican}.

Despite these advances, few have investigated whether viewing trends follow particular directional patterns across countries. 
In particular, it remains unclear whether the popularity of a show in one country can predict or reflect its success in others.
Understanding such predictive relationships is important for anticipating global content spread as well as for guiding platform strategies, informing content localization and distribution, and supporting cultural policy decisions.

To address this question, we focus on Netflix, the world’s most subscribed streaming platform. 
Its global reach, near-simultaneous release of content, and standardized popularity rankings make it a particularly suitable case for examining whether the popularity of a show in one country can predict or reflect its success in others.

Statistical physics and information theory provide useful tools for studying social systems, especially in areas like opinion dynamics, content spread, and collective behavior.
These methods treat society as a network of interacting individuals and help researchers measure influence and information flow in complex systems~\cite{durlauf1999can,peng2017social}. 
Building on this perspective, statistical-physics-inspired approaches have been applied across diverse domains, including cultural diffusion through music~\cite{duenas2023structure}, the interplay between media and public attention~\cite{kwak2018we}, and the role of international trade networks in the spread of COVID-19~\cite{antonietti2022world}.
In parallel, information-theoretic approaches have also been widely used to analyze diverse types of data, such as estimating the limits of predictability in human mobility using entropy measures~\cite{song2010limits}, examining influence among news outlets and troll accounts on Twitter~\cite{south2022information}.
In this study, we analyze cross-country viewing patterns with an information-theoretic approach, based on 822 days of daily top-ranked Netflix TV shows across 71 countries.

Our analysis is based on three core measures from information theory--Shannon entropy, mutual information, and Kullback-Leibler (KL) divergence~\cite{shannon1948mathematical, cover1999elements}.
Entropy measures how varied the top-ranked shows are within a country.
Mutual information shows how similar the viewing patterns are between two countries.  
KL divergence can be used to detect temporal asymmetries and identify possible directional patterns of content popularity across countries.

Our results show three main findings.
First, content diversity varies across regions: North America and Europe exhibit highly dynamic viewing preferences, while East and Southeast Asia (ESA) display more persistent trends with long-running popular shows.
Second, mutual information reveals clear regional clusters, with ESA countries showing particularly strong alignment in their consumption patterns.
Third, our KL-based analysis identifies directional patterns of spread. We detect inter-regional flows from ESA and South America toward North America and Europe, as well as intra-regional spreads from Korea and Thailand to other ESA countries. Notably, Singapore often acts as a connector when ESA trends extend beyond the region.

Together, these findings offer new insight into how TV show popularity evolves and spreads across countries over time.
Our approach not only reveals which countries share similar viewing preferences, but also highlights which ones tend to lead and which tend to follow.
This contributes to a deeper understanding of content spread in era of global streaming platforms.
It should be noted, however, that the ``spread'' observed here does not necessarily imply direct cultural transmission between countries. 
Rather, it reflects viewing data shaped by various external factors, including Netflix’s platform policies, differences in viewing availability across regions, and broader socioeconomic conditions.

The remainder of this paper is organized as follows.
Section 2 describes the dataset and preprocessing steps.
Section 3 presents the results and analysis, with subsections on entropy, mutual information, and KL divergence.
Section 4 discusses the implications and limitations of our findings.
Finally, Section 5 concludes with a summary and possible directions for future research.

\section{Data}
We collected the daily top-ranked TV shows on Netflix across 71 countries over a period of 822 days, from July 1, 2020 to September 30, 2022.
These data were obtained from FlixPatrol, a service that tracks Netflix’s publicly available Top-10 rankings.
A total of 561 unique shows appeared in the dataset, each labeled with an index from 1 to 561 for identification.
For each country index $k$ (where $k = 1, 2, \dots, 71$), the dataset is represented by
\beqa
  S_k \a= \{ \sdk{1}{k}, \sdk{2}{k}, \dots, \sdk{822}{k} \},
\eeqa
where $\sdk{t}{k}$ is a number between 1 and 561, indicating the top-ranked shows in country~$k$ on day~$t$.
For example, $\sdk{117}{20}$ represents the top-ranked show index in the United States~($k = 20$) on October 25, 2020~($t = 117$).
On this day, $\sdk{117}{20}$ was 473, which corresponds to the show, \textit{The Queen's Gambit}.
In addition to the 71 country-level datasets, we include a global dataset $S_0$ in our analysis~\cite{global_definition}.
For more information on country indices, see Appendix~A.

\section{Results and Analysis}

\subsection{Shannon Entropy}

\begin{figure}[t]
\centering
\includegraphics[width=0.9\linewidth]{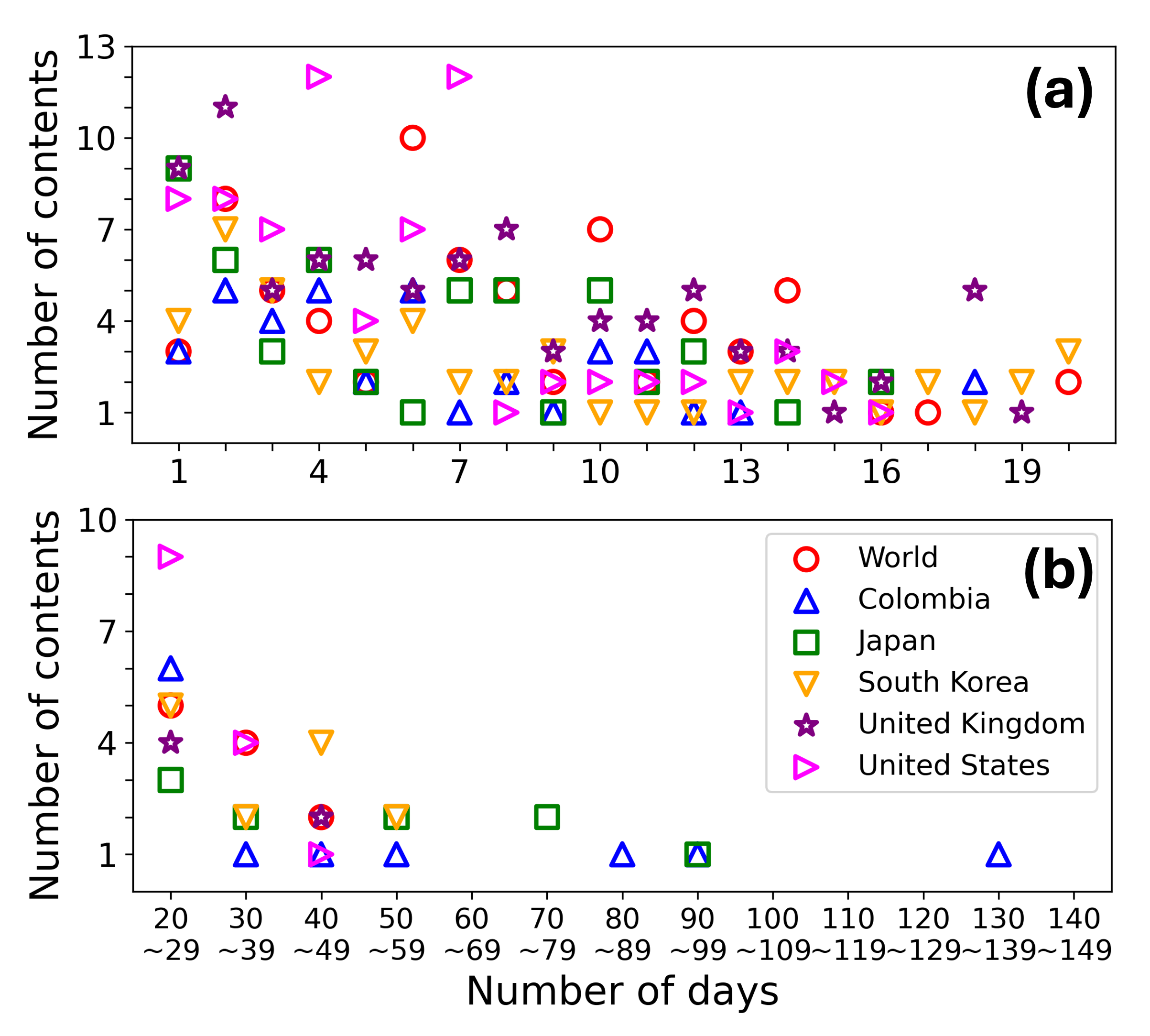}
\caption{
  Distribution of the number of distinct shows that were top-ranked over different time intervals.
  The $x$-axis indicates the total number of days (out of the 822-day observation period) that a show remained at the top rank (not necessarily consecutive), and the $y$-axis indicates how many different shows fall into that category.
  (a) presents the distribution at daily intervals from day 1 to day 20.
  (b) presents the distribution at 10-day intervals from day 20 to day 150.
  See text for details.
}
\label{fig:Figure1_Number_of_contents}
\end{figure}

Now, we examine how long each show stayed at the top rank in a given country during the 822-day observation period.
Figure~\ref{fig:Figure1_Number_of_contents} shows the frequency distribution of these durations.
On the $x$-axis, each value represents the total number of days (not necessarily consecutive) that a show remained at the top rank, while the $y$-axis indicates how many different shows fall into that category.
In other words, each point in the graph corresponds to a pair: ``duration of top ranking (days)'' on the $x$-axis and ``number of shows with that duration'' on the $y$-axis.

For example, in the United States, seven different shows held the top position for exactly three days in total (not necessarily consecutive days).
This case is indicated by a magenta left-pointing triangle located at (3, 7) on the graph.
Panel (a) of Figure~\ref{fig:Figure1_Number_of_contents} shows the distribution at daily intervals for short durations (1–20 days), while panel (b) aggregates the same information into 10-day bins for longer durations (20–150 days).
This separation allows us to highlight both the diversity of short-lived top-ranked shows and the persistence of long-running ones.

In Figure~\ref{fig:Figure1_Number_of_contents}(a), we observe that for shorter durations (i.e., a small number of days), the United States and the United Kingdom have a higher number of top-ranked shows compared to other countries.
In contrast, Figure~\ref{fig:Figure1_Number_of_contents}(b) illustrates that only Colombia and Japan have shows that remained top-ranked over 70 days.
This suggests that while the United States and the United Kingdom display more diverse viewing preferences, Colombia and Japan exhibit stronger and more sustained preferences for specific shows.

Based on this, we define the random variable $X_k$, which represents the index of the top-ranked TV show in country $k$ on a given day.
The variable $X_k$ takes values from the set of possible show indices, $\mathcal{S} = \{ 1, 2, \dots, 561 \}$, corresponding to the 561 unique shows observed in the dataset.
The probability distribution of $X_k$, denoted by $P_k$, is derived from the top-ranked TV show data $S_k$ for country $k$.
The probability data is given by
\begin{equation}
P_k = \{ \pdk{}{(k,1)}, \pdk{}{(k,2)}, \dots, \pdk{}{(k,561)} \},
\end{equation}
where $\pdk{}{(k,s)}$ represents the proportion of days that show $s$ was top-ranked in country $k$ over the 822-day period.
This value represents the empirical probability that show $s$ appears as the top-ranked show in country $k$.

\begin{figure}[!htbp]
\centering
\includegraphics[width=0.9\linewidth]{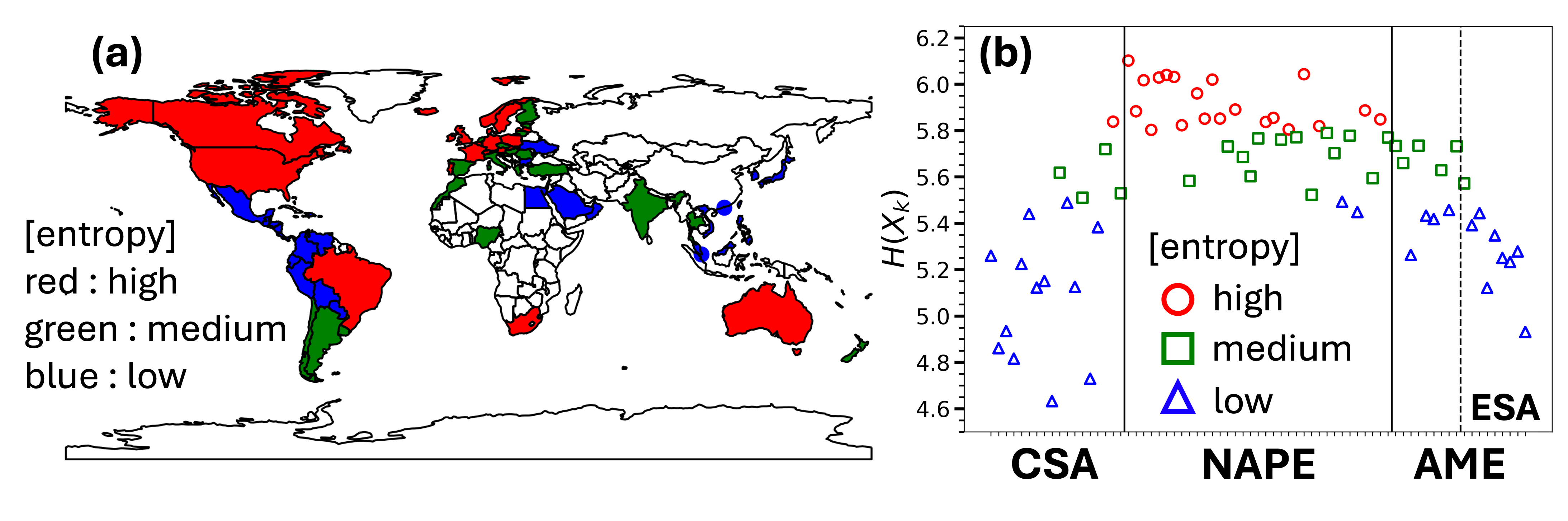}
\caption{
  (a) shows the grouping of countries by Shannon entropy, as defined by Eq.~(\ref{e.se}), displayed in three distinct colors. Red indicates high entropy values (above 5.8), green represents medium entropy values (5.5–5.8), and blue denotes low entropy values (below 5.5).
  (b) shows the entropy values by country, corresponding to the color coding in (a). The labels `CSA', `NAPE', and `AME' denote the `Central and South America Group', `North America and Europe Group', and `Asia and Middle East Group', respectively. Within each group, countries are ordered by longitude, as detailed in Appendix A.
  }
\label{fig:Figure2_Shannon_entropy}
\end{figure}

Shannon entropy quantifies the uncertainty or diversity of a probability distribution.
Using the probability distribution of $X_k$, the Shannon entropy $H(X_k)$ is computed as 
\begin{equation}
  H(X_k) = - \sum_{s \in \mathcal{S}} \pdk{}{(k,s)} \log_2 \pdk{}{(k,s)},
\label{e.se}  
\end{equation}
where we define $\pdk{}{(k,s)} \log_2 \pdk{}{(k,s)} = 0$ when $\pdk{}{(k,s)} = 0$.

In previous studies, countries were clustered into the `Central and South America Group (CSA)', `North America and Pan-Europe Group (NAPE)', and `Asia and Middle East Group (AME)' based on similarities in TV show consumption~\cite{lee2025global}. 
We also adopt this grouping in our analysis.
Figure~\ref{fig:Figure2_Shannon_entropy}(a) shows the classification of countries by their Shannon entropy values $H(X_k)$, divided into three ranges: red for high values (above 5.8), green for medium values (5.5–5.8), and blue for low values (below 5.5).
The values at 5.5 and 5.8 were chosen so that the countries are divided into groups of relatively comparable size.
Figure~\ref{fig:Figure2_Shannon_entropy}(b) presents the entropy values for each country, organized by the CSA, NAPE, and AME groups. The detailed ordering of countries within each group is provided in Appendix~A. 
According to this graph, NAPE countries are mostly red or green, reflecting relatively dynamic changes in top-ranked TV show trends. 
In contrast, CSA and AME countries are largely green or blue, consistent with more stable and persistent viewing patterns. 
Within AME, the `East and Southeast Asian (ESA)' subset is almost entirely in the blue category—with the exception of a single country—which constitutes a noteworthy observation.

\subsection{Mutual Information}
To analyze the relationship between content consumption patterns across countries, we use mutual information.
For this, we define the joint appearance probability $P_{k_1,k_2}$ for two countries
$k_1$ and $k_2$ as 
\begin{equation}
  P_{k_1,k_2} \!= \{ \pdk{}{(k_1,1;k_2,1)}, \pdk{}{(k_1,1;k_2,2)}, \dots,
  \pdk{}{(k_1,1;k_2,561)}, \pdk{}{(k_1,2;k_2,1)}, \dots, \pdk{}{(k_1,561;k_2,561)} \},
\end{equation}
where $\pdk{}{(k_1, s_1; k_2, s_2)}$ represents the proportion of days, out of the 822-day period, when TV show $s_1$ was top-ranked in country $k_1$ while show $s_2$ was top-ranked in country $k_2$ on the same day.
Using this, the mutual information $I(X_{k_1},X_{k_2})$ between countries $k_1$ and $k_2$ is calculated by
\begin{equation}
  I(X_{k_1},X_{k_2})
  = \sum_{s_1 \in \mathcal{S}} \sum_{s_2 \in \mathcal{S}} \pdk{}{(k_1,s_1;k_2,s_2)}
  \log_2 \left( \frac{\pdk{}{(k_1,s_1;k_2,s_2)}}{\pdk{}{(k_1,s_1)} \pdk{}{(k_2,s_2)}} \right).
\end{equation}
In this formula, if $\pdk{}{(k_1, s_1; k_2, s_2)} = 0$, the corresponding term in the sum is defined to be zero.

\begin{figure}[t]
\centering
\includegraphics[width=0.9\linewidth]{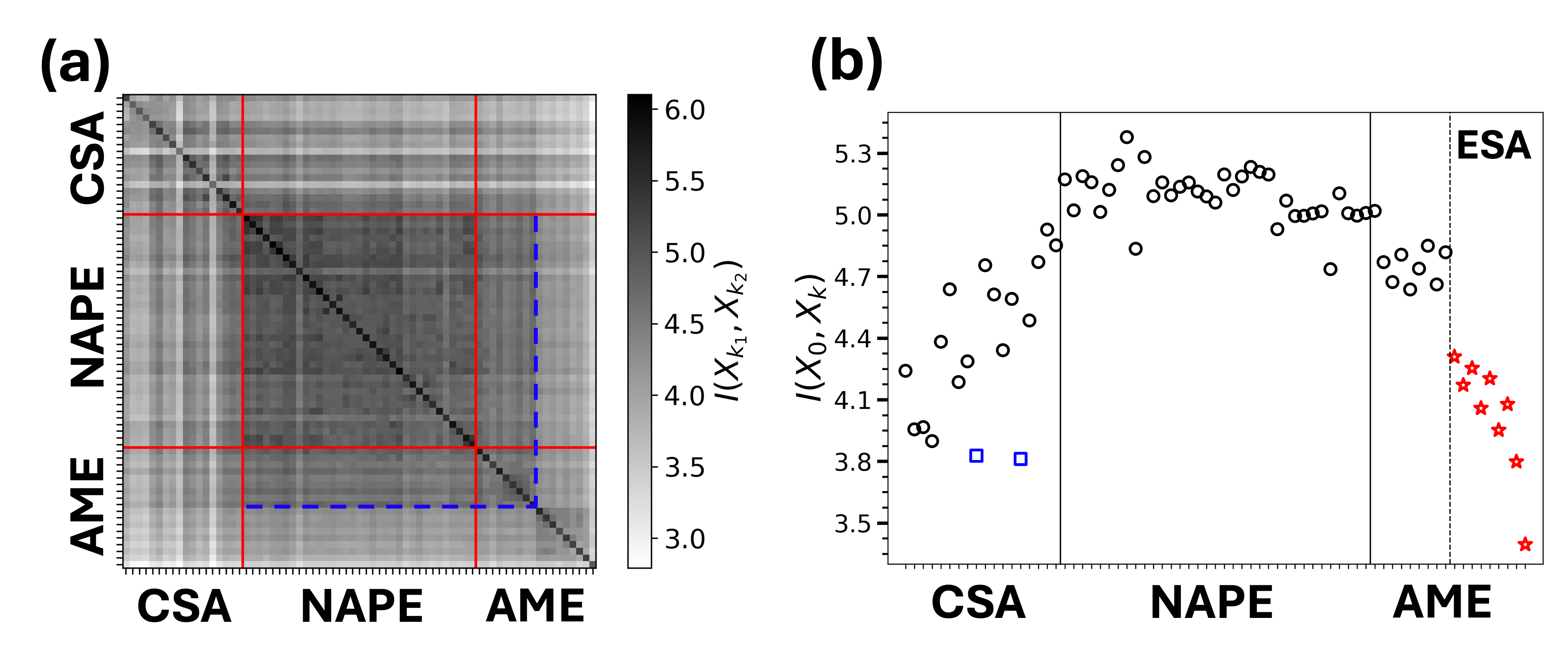}
\caption{
(a) shows the mutual information between countries, represented by grayscale shading where darker shades indicate higher values.
(b) presents the mutual information between the global dataset and each country.
The labels `CSA', `NAPE', and `AME' denote the `Central and South America Group', `North America and Europe Group', and `Asia and Middle East Group', respectively.
Countries are ordered by longitude within each group, as detailed in Appendix A.
The blue dashed line in (a) marks the boundary dividing the AME group, with western AME countries sharing higher mutual information with NAPE countries compared to East and Southeast Asia (ESA).
In (b), red stars highlight ESA countries with low mutual information with the global dataset.
Blue squares indicate Colombia and Bolivia, which also exhibit low mutual information with the global dataset.
}
\label{fig:Figure3_Mutual_information}
\end{figure}

Figure~\ref{fig:Figure3_Mutual_information}(a) presents the mutual information between countries in the form of a symmetric matrix, with countries grouped according to the classification from Lee et al.\cite{lee2025global}.
Additionally, Figure~\ref{fig:Figure3_Mutual_information}(b) shows the mutual information between each country’s dataset $S_k$ and the global dataset $S_0$, also organized by group.
The ordering of countries within each group is based on longitude, as detailed in Appendix~A.

As shown in Figure~\ref{fig:Figure3_Mutual_information}(a), the mutual information matrix reveals two prominent dark blocks, one large block in the center and a smaller one at the bottom right.
This pattern appears when countries are grouped into CSA, NAPE, and AME, and ordered by longitude within each group.
The central block does not correspond solely to the NAPE group.
Instead, it includes all NAPE countries along with several CSA and AME countries that are geographically close to NAPE.
This suggests that countries in the NAPE group not only share high internal mutual information but also exhibit overlapping TV show trends with the Middle East (part of AME) and certain CSA countries, such as Brazil, Chile, and Uruguay.
However, Colombia and Bolivia appear to be exceptions, as indicated by the two bright horizontal and vertical lines within the CSA region in the figure.
In contrast, the right side of the AME group, which contains ESA countries, shows weak connections with other countries but strong internal mutual information, suggesting a distinct and cohesive pattern in top-ranked TV show preferences.

A similar pattern appears in Figure~\ref{fig:Figure3_Mutual_information}(b), which illustrates the mutual information between each country’s dataset $S_k$ and the global dataset $S_0$.
Countries in the central block show high mutual information with the global dataset, indicating that their TV show preferences align more closely with global trends.
In contrast, ESA countries (marked with red stars) exhibit lower mutual information with the global dataset, suggesting more independent viewing patterns.
Furthermore, Colombia and Bolivia (marked with blue squares) in the CSA group also show low mutual information with the global dataset, indicating distinct preferences for TV shows. 

The previous study grouped countries based on Netflix’s top 10 rankings~\cite{lee2025global}, whereas this study focuses solely on top-ranked shows.
In the grouping with top 10 rankings, Middle Eastern countries clustered with Asia, but the current analysis shows them aligning more closely with the NAPE group.
This may suggest that Middle Eastern countries share top-watched shows with global trends but align with Asia for less popular shows.

\subsection{Kullback-Leibler (KL) Divergence}
To analyze the temporal flow of TV show preferences across countries, one might consider using transfer entropy~(TE)\cite{schreiber2000measuring}, a widely used information-theoretic measure of directional dependence between time series.
However, we chose not to apply TE for several reasons.
First, TE can misidentify indirect influences as direct ones.
For example, if countries A and B both influence country C, TE may falsely infer a direct link between A and B.
This limitation has been noted in previous studies on TE's sensitivity to common drivers\cite{runge2012escaping, bossomaier2016transfer}.
Additionally, our random variable $X_k$ can take 561 different values, leading to a combinatorial explosion in the number of possible state transitions.
Estimating TE in such a high-dimensional space becomes computationally infeasible~\cite{garland2016review}.
More importantly, our dataset includes only 822 days, which is too sparse to reliably estimate the joint and conditional probability distributions required for TE.

Beyond these technical challenges, TE has a fundamental limitation when applied to our dataset $S_k$ in the context of cross-country relationships on TV show preferences.
TE focuses only on the similarity in the pattern of changes in top-ranked shows over time between two countries. 
It does not take into account whether the shows themselves are the same.
For instance, suppose that in country $A$, the top-ranked show changes every two days:
show $\sab{A1}$ on Monday–Tuesday, $\sab{A2}$ on Wednesday–Thursday, and $\sab{A3}$ on Friday–Saturday.
In country $B$, the top-ranked show also changes every two days but with a one-day delay: $\sab{B1}$ on Tuesday–Wednesday, $\sab{B2}$ on Thursday–Friday, and $\sab{B3}$ on Saturday–Sunday.
Although the temporal pattern of change (every two days) is similar between the two countries, this does not imply that trends in country $A$ were aligned with those in country $B$ unless the shows themselves are the same.
If $\sab{B1} = \sab{A1}$, $\sab{B2} = \sab{A2}$, and $\sab{B3} = \sab{A3}$, this would suggest that TV show trends in country $A$ may have been transmitted to those in country $B$.
However, if $\sab{B1}, \sab{B2}, \sab{B3}$ are entirely different from $\sab{A1}, \sab{A2}, \sab{A3}$, then it would be difficult to argue that a cross-country relationship occurred.
Transfer entropy, however, does not take the identity of the shows into account.
It assigns the same value as long as the pattern of change is similar, regardless of whether the actual shows match.

Given these challenges, we adopt an alternative approach to examine temporal dynamics.
Instead of directly measuring information transfer, we assess whether the past content trends in country $k_2$ are similar to the current trends in country $k_1$.
To quantify this directional alignment, we use Kullback-Leibler (KL) divergence~\cite{kullback1951information}, as a measure of the difference between probability distributions.

While mutual information measures the amount of shared information between two distributions, it does not indicate which distribution better explains the other.
In contrast, KL divergence quantifies how well one probability distribution approximates another, making it more appropriate for our analysis.
Specifically, we compare the KL divergence between the past distribution of top-ranked TV shows in country $k_2$ and the current distribution in country $k_1$, and vice versa. 
That is, we assess how well the past trends in country $k_2$ explain the current trends in country $k_1$, compared to how well the past trends in country $k_1$ explain the current trends in country $k_2$. 
By comparing these two directional divergences, we aim to infer the likely direction of cross-country patterns between the two countries.
Through this approach, we are able to capture the temporal structure of content spread in TV show popularity across countries, without imposing an explicit causal model or assuming direct transmission mechanisms.

To perform this analysis, we divide the top-ranked TV show data of country $k$ into one-week periods.
This segmentation is defined by
\begin{equation}
  \Sdk{d}{k}=
  \{ \sdk{d}{k}, \sdk{d+1}{k}, \dots, \sdk{d+6}{k} \}.
\end{equation}
In this expression, $\Sdk{d}{k}$ denotes the list of top-ranked TV shows in country $k$ during the one-week period beginning on day $d$.
Based on this weekly data, we define the random variable $\Xdk{d}{k}$, which represents the top-ranked TV show in country $k$ over the same one-week period.
The random variable $\Xdk{d}{k}$ takes values from the set of possible shows $\mathcal{S} = \{1, 2, \dots, 561\}$, where each value corresponds to a specific show.

The probability distribution of $\Xdk{d}{k}$, denoted by $\Pdk{d}{k}$, is given by
\begin{equation}
\Pdk{d}{k} = \{ {\pdk{d_{}}{k}(1), \pdk{d_{}}{k}(2), \dots, \pdk{d_{}}{k}(561)} \}
\end{equation}
where $\pdk{d_{}}{k}(s)$ represents the proportion of days within the one-week period starting from day $d$ on which show $s$ appeared as the top-ranked show in country $k$.
This probability is calculated using the data in $\Sdk{d}{k}$, rather than the entire 822-day dataset.
Thus, $\pdk{d_{}}{k}(s)$ reflects the weekly appearance probability of TV show $s$ in the distribution of $\Xdk{d}{k}$.

Using the Kullback–Leibler (KL) divergence, we measure how much additional information is required when using the probability distribution of country $k_2$ at time $d_2$ to describe the probability distribution of country $k_1$ at time $d_1$.
The KL divergence is defined by
\begin{equation}
 D_\text{KL}\pa{\Pdk{d_1\!}{k_1} \parallel \Pdk{d_2\!}{k_2}}
 = \sum_{s \in \mathcal{S}} \pdk{d_1}{k_1}(s)
 \log_2 \pa{ \frac{ \pdk{d_1}{k_1}(s) }{ \ppdk{d_2}{k_2}(s) } },
\label{e.dkl}
\end{equation}
where we introduced a smoothed distribution $\ppdk{d_2}{k_2} = \pa{\pdk{d_2}{k_2} + \ \alpha}/\pa{T + \alpha V}$
of $\pdk{d_2}{k_2}$ to prevent infinite divergence when $\pdk{d_2}{k_2}(s) = 0$,
using appropriate smoothing parameter~\cite{witten2005data}.
We set $T=7$, the length of dataset $\Sdk{d}{k}$, $\alpha = \frac{1}{822}$, the inverse of the total number of days, and $V=561$, the total number of top TV shows.

The KL divergence, $D_\text{KL}(P \parallel Q)$, is an asymmetric measure that quantifies the dissimilarity between two probability distributions $P$ and $Q$.
A higher value of $D_\text{KL}(P \parallel Q)$ indicates that $Q$ is a poor approximation of $P$, meaning more information is required to describe $P$ using $Q$.
Conversely, a lower KL divergence implies that $Q$ more closely resembles $P$.
In our analysis, the KL divergence in Eq.~(\ref{e.dkl}) allows us to evaluate how well the top-ranked TV show distribution in country $k_2$ from day $d_2$ approximates that in country $k_1$ from day $d_1$.

\begin{figure}[!htbp]
\includegraphics[width=0.9\linewidth]{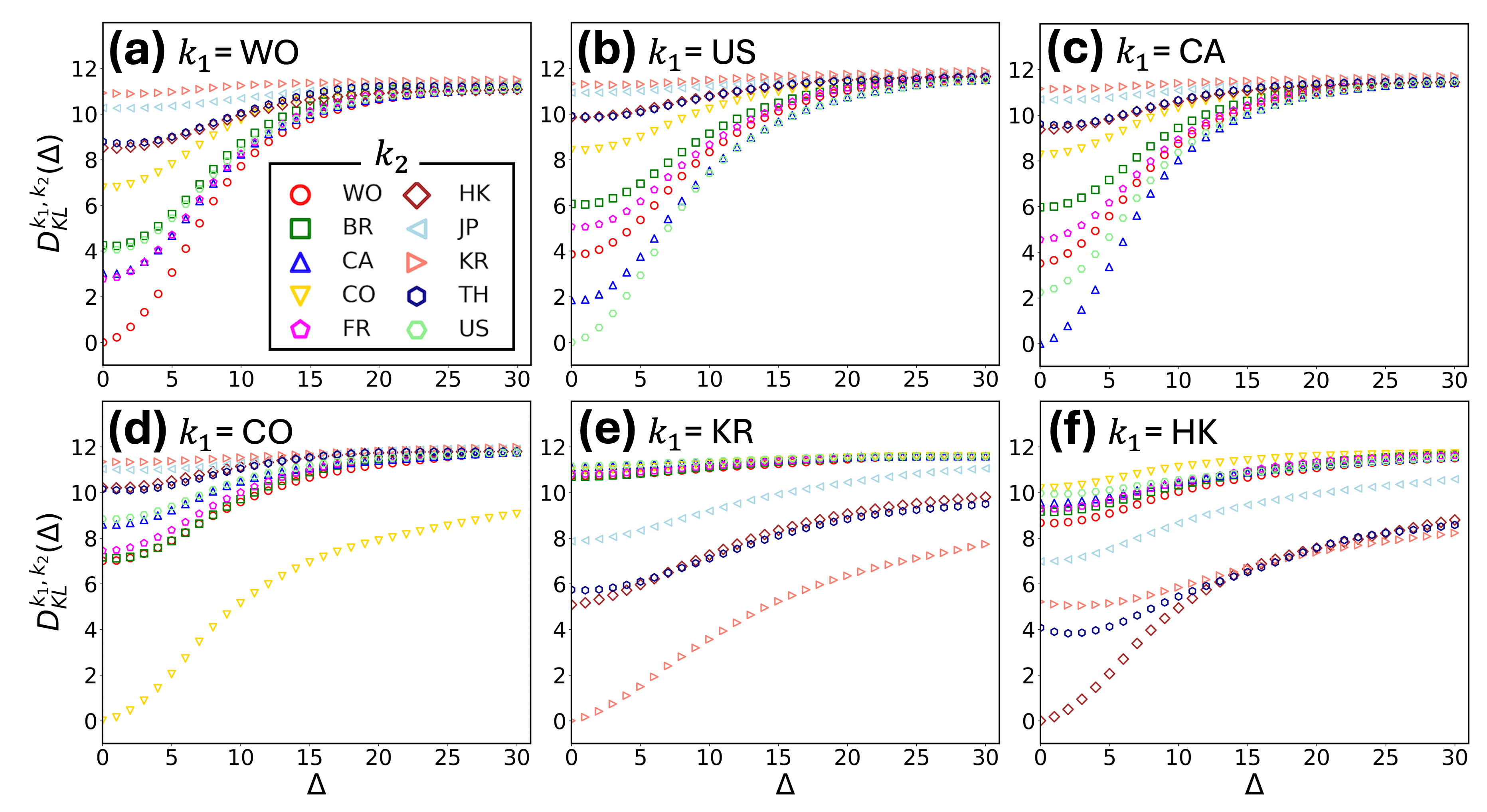}
\caption{
Average KL divergence $D_\text{KL}^{k_1,k_2}(\Delta)$ between country pairs $k_1$ and $k_2$, with a time shift $\Delta$ from 0 to 30 days. The pairs are selected from 10 representative countries among the 72 analyzed: `WO' (World), `BR' (Brazil), `CA' (Canada), `CO' (Colombia), `FR' (France), `HK' (Hong Kong), `JP' (Japan), `KR' (Korea), `TH' (Thailand), and `US' (United States).
}
\label{fig:Fig4_KL_divergence}
\end{figure}

Figure~\ref{fig:Fig4_KL_divergence} presents the average KL divergence, $D_\text{KL}^{k_1,k_2}(\Delta)$, between two countries $k_1$ and $k_2$ with a time shift of $\Delta$ days.
The average is taken over the interval from $d_{\text{start}} = 31$ to $d_{\text{end}} = 822$ and is given by
\beqa
D_\text{KL}^{k_1,k_2}(\Delta)
\a= \overline{D_\text{KL}\pa{\Pdk{d}{k_1} \parallel \Pdk{d-\Delta}{k_2}}} \nonumber\\
\a= \frac{1}{T} \sum_{d=d_{\text{start}}}^{d_{\text{end}}}
D_\text{KL}\pa{\Pdk{d}{k_1} \parallel \Pdk{d-\Delta}{k_2}},
\eeqa
where $T = d_{\text{end}} - d_{\text{start}} + 1$, and the time shift $\Delta$ ranges from $0$ to $30$.
In our dataset, a country has on average 1.83 unique top-ranked shows per 7-day window.
For example, suppose that in country $k_1$, show A was the top-ranked show for 4 days and show B for 3 days within a 7-day window.
If country $k_2$ does not include either A or B in its top-ranked shows during that 7-day window, then the KL divergence $D_\text{KL}(P_{k_1} \parallel P_{k_2})$ equals 11.64. 
When B appears once or twice in $k_2$'s distribution—while all other days still feature shows other than A or B—the divergence drops to 7.48 and 7.06, respectively.  
Similarly, if only A appears once or twice, with the rest of the days still occupied by non-A/B shows, the divergence is 6.10 and 5.53, respectively.  
Finally, when both A and B appear once each and the remaining five days are filled with different shows, the divergence further drops to 1.95.
These baseline cases serve as a reference for interpreting the actual values in Fig.~\ref{fig:Fig4_KL_divergence}, which are discussed below.

Let us first examine the cases where $k_1 = k_2$ in Figure~\ref{fig:Fig4_KL_divergence}.  
When $\Delta = 0$, the two datasets are identical, and thus $D_\text{KL}^{k_1,k_2}(0) = 0$.  
As $\Delta$ increases, the KL divergence also increases, but the rate and shape of this increase differ across countries.  
In the cases of World (panel a), the United States (panel b), and Canada (panel c), the divergence rises rapidly with increasing $\Delta$, reaching values close to 11 when $\Delta = 30$.
These values correspond to a scenario where there is virtually no overlap in the top-ranked shows when the target country $k_1$’s distribution consists of show A top-ranked for 4 days and show B top-ranked for 3 days.
In contrast, for Colombia (panel d), South Korea (panel e), and Hong Kong (panel f), the increase is more gradual, and the values remain relatively low even at $\Delta = 30$.  
This suggests that content trends in the former three cases change more quickly, whereas in the latter three countries, some top-ranked shows remain dominant even a month later.  
This pattern is consistent with the results observed earlier in Fig.~\ref{fig:Figure2_Shannon_entropy}.

For cases where $k_1 \ne k_2$, the KL divergence generally increases with $\Delta$, indicating that the dissimilarity between the distributions grows as the time gap widens.
However, there are exceptions.
In some pairs of $k_1$ and $k_2$, the KL divergence is already close to the maximum value for the target country $k_1$'s distribution, even when $\Delta=0$.
For example, in panel (a) for $k_1 =$ World, $D_\text{KL}^{k_1,k_2}(0)$ with $k_2 =$ Korea or Japan exceeds 10, suggesting that the top-ranked shows in Korea and Japan are almost completely independent of the world’s top-ranked shows.
Additionally, there are cases where $\Delta = 0$ does not yield the minimum KL divergence.
In panel (f), where $k_1$ is Hong Kong, the curves for $k_2 =$ Thailand and South Korea show their minimum divergence at a nonzero $\Delta$.
This indicates that the past distributions of Thailand and South Korea better explain the present state of Hong Kong than their concurrent distributions, suggesting a potential directional patterns from these countries to Hong Kong.

Next, we introduce a threshold value $D_\text{th}$ for the KL divergence and compare the range of $\Delta$ over which the divergence remains below this threshold, that is, $D_\text{KL}^{k_1,k_2}(\Delta) < D_\text{th}$.
We define $\delm$ as
\beqa
  \delm
  \a= \operatorname*{argmax}_{\Delta}  \ D_\text{KL}^{k_1,k_2}(\Delta) < D_\text{th},
\eeqa
when there exists at least one $\Delta$ in the range $0 \le \Delta \le 30$ satisfying the condition.
Otherwise, $\delm$ is not defined.

For example, consider Fig.~\ref{fig:Fig4_KL_divergence}(e),
where $k_1$ represents South Korea.  
When the threshold is set to $D_\text{th} = 6$, the KL divergence from Hong Kong remains below this value for up to 5 days.  
Thus, we have $\delme{70}{67}{6} = 5$, where South Korea (SK) has index 70 and Hong Kong (HK) has index 67, as listed in Appendix~A.  
In contrast, in panel (f), where $k_1$ is Hong Kong, the KL divergence from South Korea stays below the threshold for up to 10 days, hence, $\delme{67}{70}{6} = 10$.
This indicates that, under the threshold $D_\text{th} = 6$, South Korea’s past show distribution explains Hong Kong’s current trends over a longer historical window than Hong Kong’s past explains South Korea’s ($\delme{67}{70}{6} > \delme{70}{67}{6}$).

\begin{figure}[t]
\includegraphics[width=0.9\linewidth]{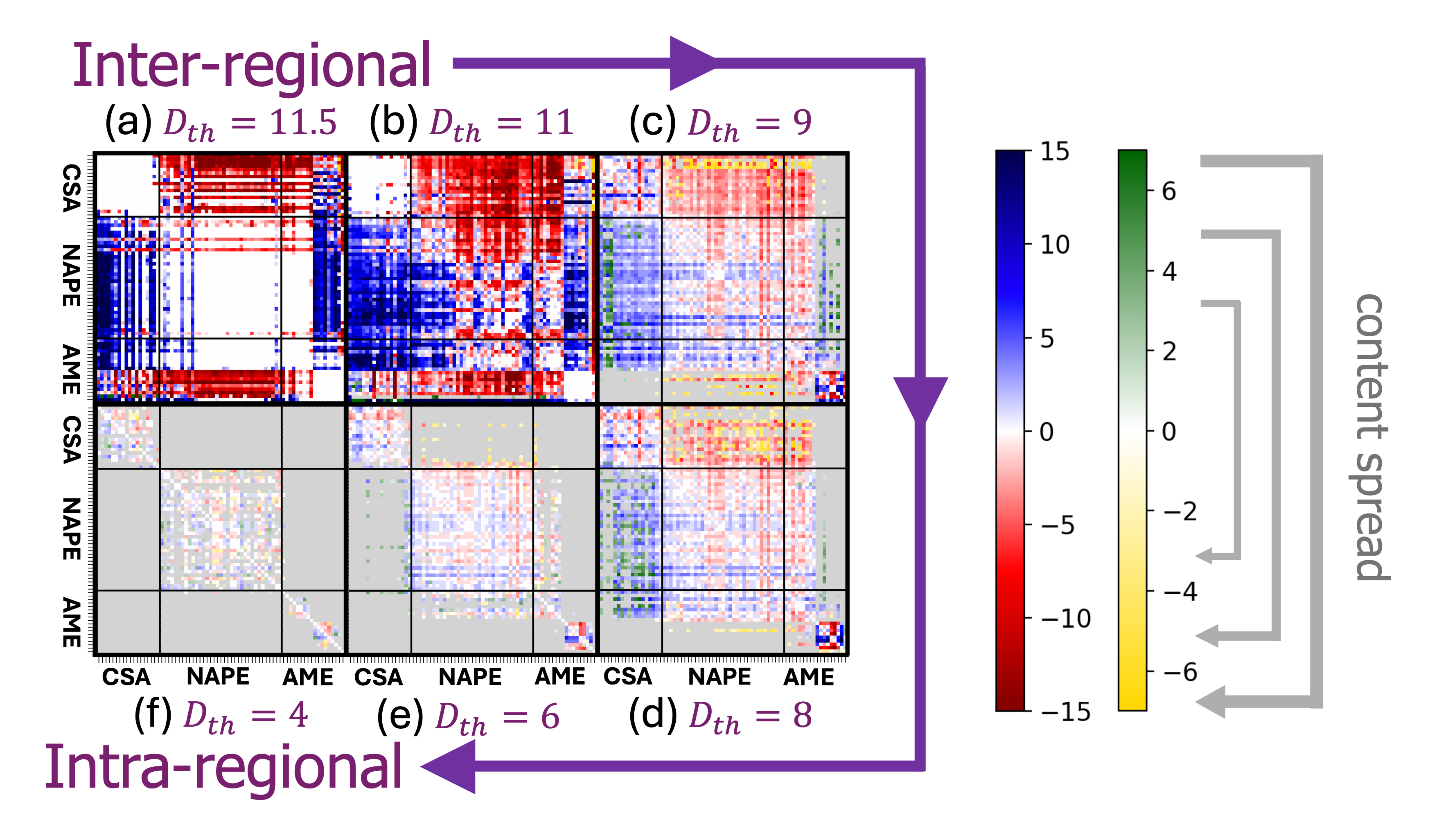}
\caption{
Comparison of $\deld$ values defined in Eq.~(\ref{e.deld}) between country pairs.
Each cell shows how well the top-ranked show list of the column country ($k_2$) explains that of the row country ($k_1$), relative to the reverse.
Threshold values $D_\text{th}$ are set to (a) 11.5, (b) 11, (c) 9, (d) 8, (e) 6, and (f) 4.
The group labels `NAPE', `AME', and `CSA' refer to the `North America and Pan-Europe Group', `Asia and Middle East Group', and `Central and South America Group', respectively.
Countries within each group are ordered as listed in Appendix A.
See main text for details on the color scheme and interpretation.
}
\label{fig:Fig5_KL_divergence_table}
\end{figure}

We perform this comparison for all country pairs and summarize the results in   Figure~\ref{fig:Fig5_KL_divergence_table}.  
For a given threshold $D_\text{th}$, we define the asymmetry $\deld$ as the difference in the maximum time window $\Delta_\text{max}$ between two countries,  
\beqa
  \deld  
  \a= \delm - \delmb,
\label{e.deld}  
\eeqa  
when both $\delm$ and $\delmb$ are defined.  
This value captures how much longer the past data of one country explains the present state of another country, compared to the reverse direction.

We visualize the values of $\deld$ using color in Figure~\ref{fig:Fig5_KL_divergence_table}, for six different threshold values $D_\text{th}$: (a) 11.5, (b) 11, (c) 9, (d) 8, (e) 6, and (f) 4.
We vary the threshold because different threshold levels allow us to detect different levels of closeness.
When the top-ranked show lists of countries $k_1$ and $k_2$ are very different, the KL divergence $D_\text{KL}^{k_1,k_2}(\Delta)$ tends to be large, meaning more information is needed for $k_2$ to predict $k_1$’s trends.
In such cases, we must use a large threshold $D_\text{th}$ to detect any subtle similarity.

In contrast, when the show lists of $k_1$ and $k_2$ are similar, their KL divergence values are generally small.
If we use a large threshold in this case, the divergence may remain below the threshold even at $\Delta = 30$, the maximum time lag we consider. 
To capture the direction and strength of information pattern between such closely related countries, we need to use a smaller threshold value.

If $D_{\text{th}}$ is set too large, countries can trivially explain each other’s distributions, resulting in no meaningful directional patterns. Conversely, if $D_{\text{th}}$ is set too small, countries cannot explain each other at all and the links vanish. We therefore focus on the interval $4 \leq D_{\text{th}} \leq 11.5$, where directional patterns remain interpretable and non-trivial.

In Figure~\ref{fig:Fig5_KL_divergence_table}, rows and columns represent countries, grouped under the labels `CSA' (Central and South America Group), `NAPE' (North America and Pan-Europe Group), and `AME' (Asia and Middle East Group), with the ordering of countries within each group provided in Appendix~A.
This figure shows, for a given threshold, whether the column country can explain the row country over a longer historical window than in the reverse direction.

The values of $\deld$ are shown using a red-to-blue color scheme, where color is applied only when both $\delm$ and $\delmb$ are defined.
A separate color is used when either $\delm$ or $\delmb$ is not defined, as explained below.
Bluer (more positive) values indicate that the column country can explain the row country further back in time, while redder (more negative) values indicate the opposite.

For example, with a threshold $D_\text{th} = 6$ as discussed earlier, South Korea can explain Hong Kong's distribution up to 5 days further into the past than Hong Kong can explain South Korea’s.
Accordingly, we have $\delde{67}{70}{6} = 5$ (blue) and $\delde{70}{67}{6} = -5$ (red), meaning the cell with South Korea as the column and Hong Kong as the row is marked as $+5$, and the reverse cell as $-5$.
If both countries can explain each other equally far back in time under the given threshold, the corresponding cells are shown in white.
For instance, if South Korea’s show list from up to 3 days ago can explain Hong Kong’s current distribution, and vice versa, ($\delme{70}{67}{D_\text{th}} = 3$, $\delme{67}{70}{D_\text{th}} = 3$), then both cells are marked as white, meaning $\delde{70}{67}{D_\text{th}} = 0$ and $\delde{67}{70}{D_\text{th}} = 0$.
This indicates that there is no directional asymmetry in explanatory power under the given threshold.

When only one of $\delm$ or $\delmb$ is defined, we use a yellow-to-green color scheme to represent the value.
If only $\delm$ is defined, we display its value in green.
Conversely, if only $\delmb$ is defined, we show $-\delmb$ in yellow.
Greener (more positive) values indicate that the column country ($k_2$) can explain the row country ($k_1$) further back in time, while yellower (more negative) values indicate that the row country can explain the column country further back in time.
These yellow–green cases may suggest an even stronger directional tendency than the red–blue pairs, as one direction lacks any explanatory power under the threshold, while the other retains some.

There are also cases where neither $\delm$ nor $\delmb$ is defined.
This occurs when both
$D_\text{KL}^{k_1,k_2}(\Delta)$ and
$D_\text{KL}^{k_2,k_1}(\Delta)$
remain above the threshold $D_\text{th}$ for the entire range $0 \le \Delta \le 30$.
In such cases, both the $(k_1, k_2)$ and $(k_2, k_1)$ cells are shown in gray.

For the highest threshold value, $D_\text{th} = 11.5$, shown in panel (a), $\deld$ primarily captures differences in explanatory power between countries from different groups.
Most intra-group pairs appear white, indicating that the top-ranked show lists are similar enough for each country to explain the other equally well under this generous threshold.
This suggests strong regional similarity in content consumption patterns.
When a large amount of information difference is allowed, there is no distinct directional asymmetry within groups for most cases. 
However, some countries in the AME group, particularly those in the Middle East and Western Asia, show greater similarity to NAPE countries, consistent with the patterns observed in Figure~\ref{fig:Figure3_Mutual_information}.

While the analysis in Figure~\ref{fig:Figure3_Mutual_information} identified regional groupings based on mutual information computed at the same time point, the current approach incorporates temporal delays to capture the directional pattern of content trends across countries. 
In panel (a), we observe blue-shaded patterns from ESA (the right side of AME) countries and from the CSA group toward the left side of AME and NAPE. 
When content preferences between two countries are highly different, even weak signals, such as the delayed and occasional appearance of content from ESA or the CSA group in NAPE countries, can have a disproportionately large effect in this analysis.
Conversely, since NAPE countries tend to be closely aligned with global trends, their contribution to other regions likely occurs in a more simultaneous and widespread manner. 
As a result, their outbound signals are less detectable in this time-delay-based framework, which is better suited for identifying sequential spread.  
In this analysis, the NAPE group therefore tends to appear more as a receiver than a sender of global content dynamics.
However, this tendency may also reflect platform-level release policies or global distribution strategies, and thus should be interpreted with caution.

In panel (b), where $D_\text{th} = 11$, slightly lower than in panel (a), directional patterns within the NAPE group begin to emerge.
As noted earlier, NAPE countries are generally aligned with global trends, making it difficult to detect directional patterns from them to other groups.
However, at this threshold, sequential dynamics within the NAPE group itself become visible.
In particular, we observe a pattern from North America and Western Europe (on the left side of the NAPE group) toward Eastern Europe (on the right side of the NAPE group) and toward the Middle East and Western Asia (on the left side of the AME group).
In contrast, no strong directional patterns are yet observed within ESA or within the CSA group.
This may be because countries in these groups still show relatively similar content consumption patterns, allowing them to explain each other to a comparable degree under the current threshold.

In panel (c), where $D_\text{th} = 9$, directional patterns start to emerge within ESA (the right side of the AME group) and within the CSA group.
Additionally, we observe yellow-to-green color pairs appearing between ESA and other regions.
The presence of green cells in the ESA columns indicates a directional tendency from ESA toward West Asia and NAPE.
Moreover, the rapid disappearance of interregional connections involving ESA indicates that it maintains relatively independent content consumption patterns, distinct from those of other regions.

As the threshold decreases (panels (d), (e), and (f)), cross-group connections progressively vanish (gray cells), and even directional patterns within regions become less pronounced.
These patterns suggest that under stricter information constraints, popular TV-show viewing patterns become more localized, with each region showing increasingly independent behavior and fewer detectable connections across regional boundaries.

To further check the stability of our results, we examined whether the signs of directional links change when varying $D_{\text{th}}$ from 0 to 12. We found that more than 80\% of all pairs retain identical signs (ignoring zeros), suggesting that the overall directional patterns are largely insensitive to the choice of $D_{\text{th}}$.

\begin{figure}[!htbp]
\includegraphics[width=0.9\linewidth]{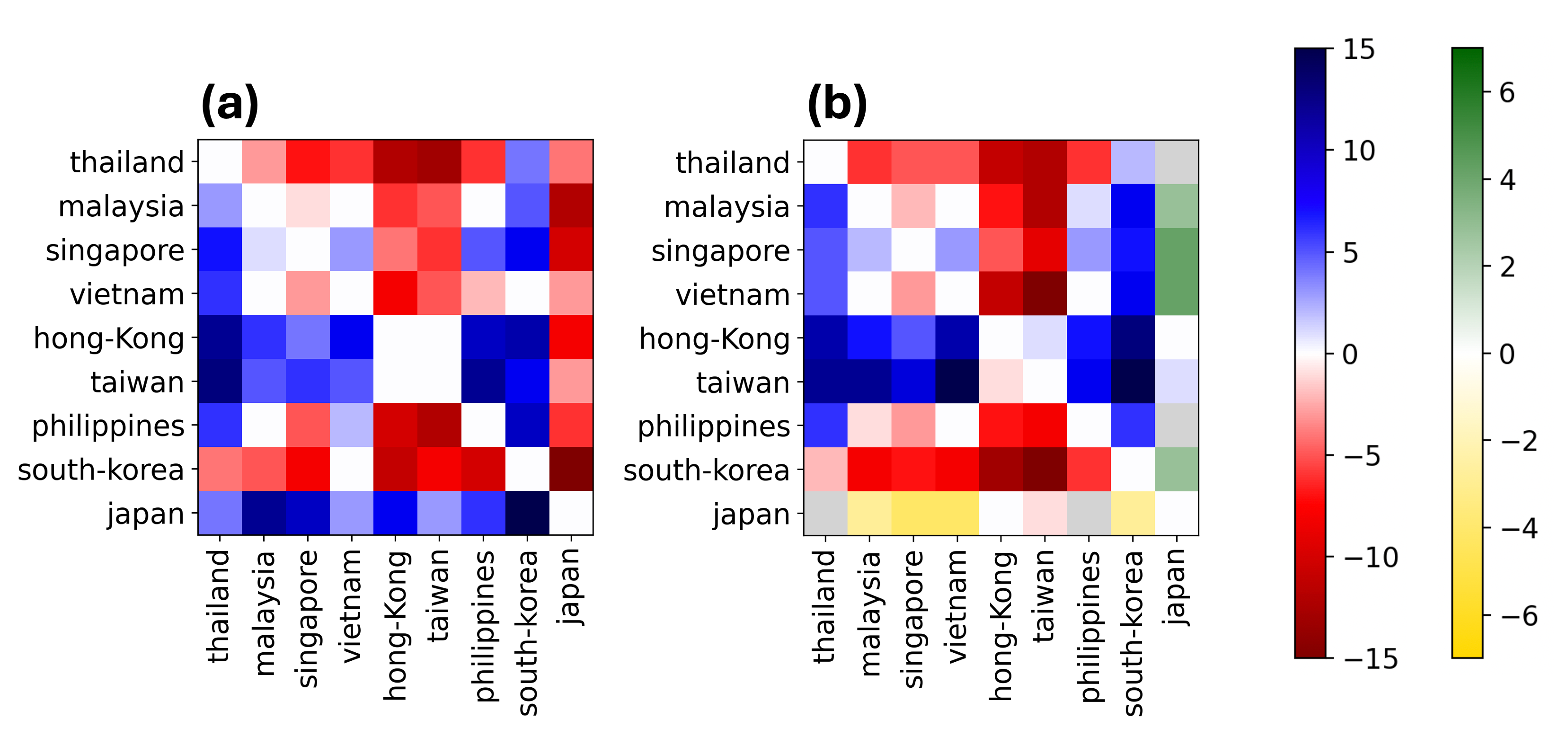}
\caption{
  Enlarged view of the intra-group comparison from Figure~\ref{fig:Fig5_KL_divergence_table}, focusing on ESA countries.
  Panels (a) and (b) show results for threshold values $D_\text{th} = 9$ and $D_\text{th} = 8$, respectively.
}
\label{fig:Fig6_KL_divergence_table_asia_within}
\end{figure}

Figure~\ref{fig:Fig6_KL_divergence_table_asia_within} presents a magnified view of ESA from Figure~\ref{fig:Fig5_KL_divergence_table}, focusing on intra-regional patterns.
Panels (a) and (b) correspond to threshold values $D_\text{th}$ of 9 and 8, respectively.

In panel (a), South Korea and Thailand tend to produce blue cells when they appear as column countries, suggesting that they act as sources of TV-show popularity within the region.
In contrast, Hong Kong, Taiwan, and Japan tend to produce red cells in the column position, indicating more receiver-like characteristics.
A clear directional pattern is observed from South Korea to Thailand, suggesting that South Korea may serve as an initial source of TV-show trends in ESA.

In panel (b), a similar pattern is observed, but under the stricter threshold, Japan shifts slightly toward a sender role, as indicated by the appearance of green cells when it serves as a column country.
This suggests that although Japan's data may still explain trends in other countries, the reverse is no longer true at this threshold level.
This asymmetry implies that Japan's content consumption patterns are structurally independent from those of its neighbors, making them difficult to predict based on regional data.
Nevertheless, the fact that Japan retains explanatory power for other countries indicates that some TV-show trends may originate in Japan and spread to other parts of the region.

\begin{figure}[!htbp]
\includegraphics[width=0.9\linewidth]{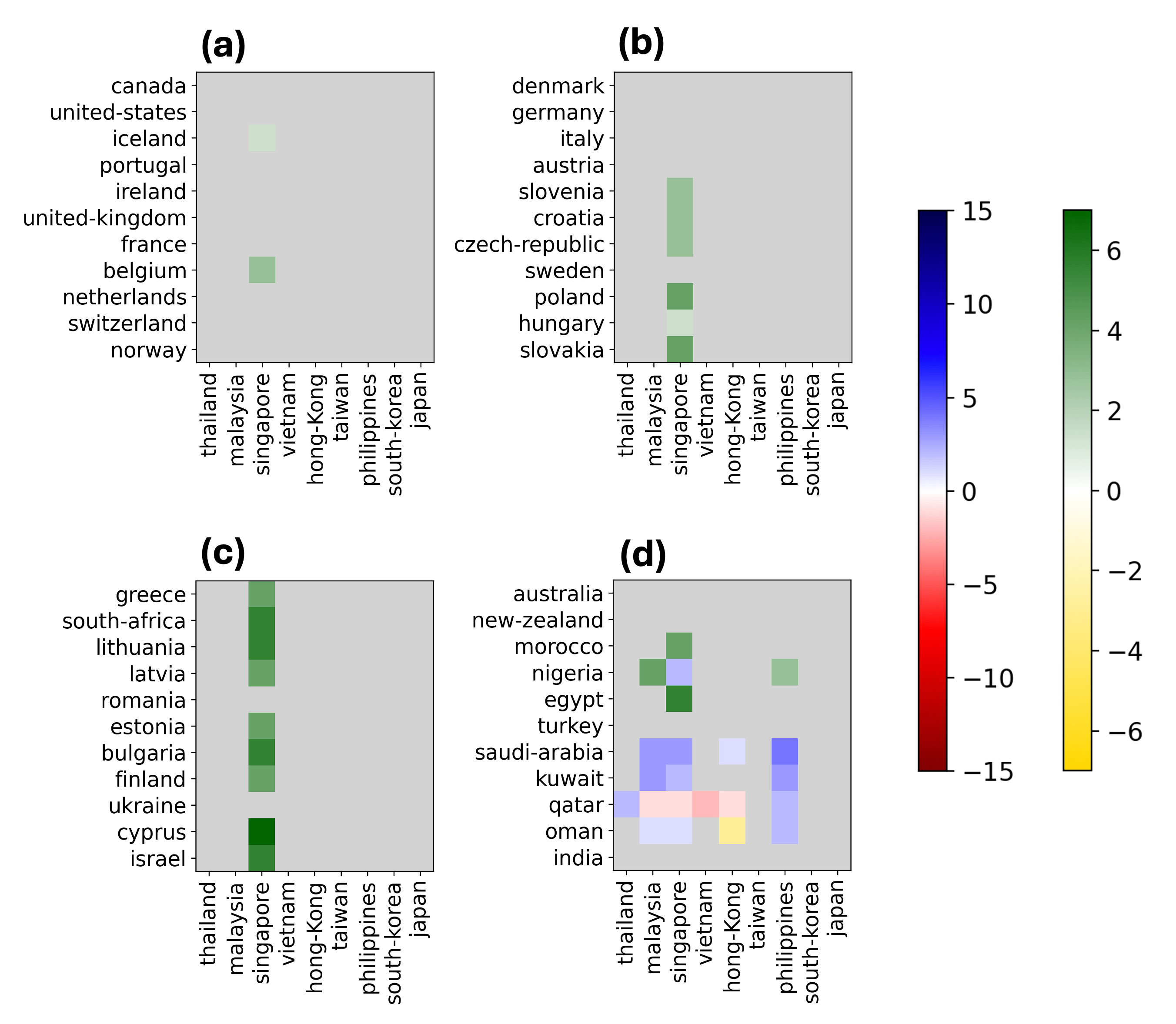}
\caption{
  Enlarged view of the between-group comparison in Figure~\ref{fig:Fig5_KL_divergence_table}, showing directional relationships from ESA countries to other regions at a threshold of 8.
}
\label{fig:Fig7_KL_divergence_table_asia_between}
\end{figure}

Figure~\ref{fig:Fig7_KL_divergence_table_asia_between} shows the connections between ESA countries and the rest of the world at a threshold of $D_\text{th} = 8$.
Most cells are gray, indicating that directional patterns have broken down.
However, a few notable exceptions appear.
Malaysia, Singapore, and the Philippines show directional links to several countries in Western Asia and the Middle East.
Among ESA countries, Singapore uniquely maintains directional connections to other regions, including parts of Europe and Africa.
This contrasts with Figure~\ref{fig:Fig6_KL_divergence_table_asia_within}, where South Korea and Thailand act as regional leaders within ESA.
In Figure~\ref{fig:Fig7_KL_divergence_table_asia_between}, Singapore stands out as a connector between ESA and the rest of the world.
This may be due to its relatively small KL divergence with other regions, suggesting that TV-show preferences in Singapore moderately overlap with global trends, allowing it to retain connections even under stricter information constraints. 
Such a role is consistent with previous studies showing that Singapore exhibits high betweenness centrality across multiple online consumption networks, reflecting its position as a multicultural society that bridges Southeast Asia and the western world~\cite{ng2023web}.

Altogether, the KL divergence–based analysis offers a clear framework for understanding how TV-show preferences evolve and spread across countries over time.
By tracking changes in the distribution of top-ranked shows, we found directional patterns that reveal both regional similarities and cross-group patterns.
It is important to note, however, that KL divergence does not establish causality.
Global trends that arise simultaneously in multiple countries may not be detected through temporal asymmetries alone.
While alignment between one country’s past and another’s present suggests that trends may have spread over time, it may not imply a direct cause-and-effect relationship.
Moreover, apparent relationships between countries may also be driven by common underlying factors, such as platform algorithms, rather than by direct links.
Nonetheless, our method highlights directional asymmetries in content similarity over time, helping to identify potential pathways through which TV-show trends may have diffused.
This approach not only reveals which countries share similar preferences, but also how these similarities may have developed, providing a temporal lens on global content spread.

\section{Discussion}
Our analysis of cross-country relationships and directional patterns is based on observed preferences for top-ranked Netflix content. 
While this approach reveals important structural patterns, it also comes with limitations. 

The first limitation concerns the data itself. 
Although Netflix is a global content provider, the number of titles available and their release dates vary across countries, and our analysis did not account for these differences. 
This is a clear factor that may reduce the accuracy of our results. 
Nevertheless, it should be noted that in a previous study using a network-based approach~\cite{lee2025global}, the main analysis used all available titles, but an additional robustness check in the appendix restricted the data to titles that simultaneously appeared in the top-10 lists of two countries on the same day. 
Under this narrower criterion, the resulting weights remained highly correlated with those from the original method, suggesting that despite potential limitations related to release times, broadly similar patterns may still be obtained.

Another aspect of the data limitation concerns genre classification.
The “show” category we examined encompasses multiple genres, and a more fine-grained analysis by genre could potentially yield additional insights. 
Addressing these differences in content availability and genre classification would allow for a more precise understanding of cross-country dynamics.

Second, this relates to the characteristics of the Netflix platform. 
Previous studies have demonstrated that linguistic preferences strongly shape how audiences choose media~\cite{ksiazek2008cultural}. 
Even in the broader digital environment, linguistic and geographic proximity better explain cross-audience overlap than structural connections such as hyperlinks or shared genres~\cite{taneja2016global}. 
By contrast, Netflix operates as a global distributor with the goal to breaking down national and cultural boundaries. 
As Park et al.~\cite{park2025social} point out, linguistic similarity still plays a stronger role for Netflix than for other platforms. 
However, Netflix simultaneously pursues strategies designed to minimize cultural specificity and maximize global universality in production~\cite{khazoom2025televisuality}. 
In addition, its recommendation algorithms and marketing strategies restructure consumption patterns around global preferences rather than national origin~\cite{neira2023standing}. 
These platform-specific strategies, together with powerful algorithms, inevitably influence national consumption patterns and may amplify or suppress emerging trends. 
Therefore, our results should be interpreted with the possibility that algorithmic interventions and Netflix’s platform-level strategies and policies have shaped the observed dynamics. 
For example, the tendency for NAPE countries to appear more as receivers in our analysis may also reflect such global release policies rather than purely audience-driven dynamics.

A third limitation concerns the treatment of time. 
Our analysis implicitly assumes a uniform temporal scale across countries, yet the lifetime of popular titles may vary significantly depending on country-specific characteristics. 
Larger countries may take longer to saturate target audiences. 
In addition, wealthier countries with greater availability of leisure time may exhibit shorter cycles of content replacement. 
This could partly explain our results. 
For example, the finding in Figure~\ref{fig:Figure1_Number_of_contents} that the United States experiences frequent changes in top-ranked content, while Colombia remains more stable, may reflect the combined effects of these country characteristics. 
This heterogeneity may also contribute to the differences in KL divergence observed between countries in Figure~\ref{fig:Fig4_KL_divergence}.
Future work should account for such heterogeneous temporal scales and consumption patterns across countries.

Despite these limitations, our study provides meaningful insights into the cross-country dynamics of Netflix content consumption and demonstrates the value of information-theoretic approaches for analyzing global media patterns. 
Future research could build on this framework by addressing differences in content availability, accounting for platform-level interventions, and incorporating heterogeneous temporal scales and genre-specific analyses. 
Such extensions would strengthen the robustness of the results and offer a more comprehensive understanding of global media consumption patterns.

\section{Conclusion}
This study examined global media consumption patterns by analyzing the top-ranked Netflix TV shows across 71 countries over 822 days using an information-theoretic framework. 
Through the combined use of Shannon entropy, mutual information, and Kullback-Leibler (KL) divergence, we quantified the diversity of content trends within each country, assessed synchronous similarities between country pairs, and uncovered temporal asymmetries indicative of directional content spread.

Our entropy analysis revealed substantial variation in the volatility of content trends across regions. 
North America and Pan-Europe exhibited high entropy, with frequent shifts in top-ranked shows, while ESA showed lower entropy, reflecting more stable and enduring content preferences. 
Mutual information uncovered strong regional clustering. 
Countries within the NAPE group shared similar consumption patterns and aligned closely with the global trend, while ESA countries displayed relatively independent and internally cohesive viewing preferences.

To capture directional dynamics, we introduced a KL-based asymmetry measure that compares weekly distributions of top-ranked shows across countries with varying time lags.
By adjusting the divergence threshold, we identified both fine-grained relationships within tightly synchronized regions and broader patterns across loosely connected areas. 
Under small threshold values, we detected intra-regional patterns, such as signals originating from Korea and Thailand to other countries in ESA. In contrast, under large thresholds, we observed inter-regional patterns from ESA and CSA toward the NAPE region. 
These findings highlight how dynamics in global viewing patterns depend on both regional proximity and the popularity strength of shows.

In summary, this work provides quantitative evidence of how digital content circulates globally, but it should also be interpreted with caution, as platform algorithms and country-specific consumption patterns may influence the observed dynamics. 
Future work should extend this framework by improving data accuracy, explicitly incorporating platform-level interventions, and accounting for heterogeneous temporal scales. 
Overall, our results contribute to a deeper understanding of not only who watches what, but also how and when preferences are observed across countries, with implications for content distribution strategies, localization strategies, and cultural policy in an increasingly global media environment.

\newpage
\appendix
\section*{Appendix A.}
\begin{table}[h]
\renewcommand{\arraystretch}{0.85}
\begin{footnotesize}
\begin{tabular}{c l l @{\hspace{1cm}} c l l}
\hline
Index & Country & Group & Index & Country & Group \\
\hline
1  & Mexico  & CSA   & 37 & Sweden  & NAPE \\
2  & Guatemala  & CSA  & 38 & Poland  & NAPE \\
3  & Honduras  & CSA    & 39 & Hungary  & NAPE \\
4  & Nicaragua  & CSA   & 40 & Slovakia & NAPE \\
5  & Costa Rica   & CSA  & 41 & Greece & NAPE \\
6  & Panama  & CSA  & 42 & South Africa & NAPE \\
7  & Ecuador  & CSA  & 43 & Lithuania & NAPE \\
8  & Peru  & CSA  & 44 & Latvia & NAPE \\
9  & Colombia  & CSA   & 45 & Romania  & NAPE \\
10 & Chile  & CSA   & 46 & Estonia & NAPE \\
11 & Dominican Republic & CSA  & 47 & Bulgaria & NAPE \\
12 & Venezuela  & CSA   & 48 & Finland & NAPE \\
13 & Argentina  & CSA   & 49 & Ukraine & NAPE \\
14 & Bolivia  & CSA   & 50 & Cyprus & NAPE \\
15 & Paraguay  & CSA  & 51 & Israel & NAPE \\
16 & Uruguay  & CSA   & 52 & Australia & NAPE \\
17 & Brazil  & CSA   & 53 & New Zealand & NAPE \\
18 & Spain  & CSA   & 54 & Morocco  & AME  \\
19 & Canada  & NAPE  & 55 & Nigeria & AME  \\
20 & United States  & NAPE & 56 & Egypt  & AME  \\
21 & Iceland  & NAPE  & 57 & Turkey  & AME  \\
22 & Portugal & NAPE  & 58 & Saudi Arabia & AME  \\
23 & Ireland  & NAPE   & 59 & Kuwait & AME  \\
24 & United Kingdom  & NAPE   & 60 & Qatar & AME  \\
25 & France  & NAPE & 61 & Oman & AME  \\
26 & Belgium  & NAPE   & 62 & India & AME  \\
27 & Netherlands  & NAPE  & 63 & Thailand & AME (ESA)  \\
28 & Switzerland  & NAPE  & 64 & Malaysia & AME (ESA)  \\
29 & Norway & NAPE & 65 & Singapore & AME (ESA)  \\
30 & Denmark  & NAPE  & 66 & Vietnam & AME (ESA)  \\
31 & Germany  & NAPE  &  67 & Hong Kong  & AME (ESA)  \\
32 & Italy  & NAPE  & 68 & Taiwan & AME (ESA)  \\
33 & Austria  & NAPE  & 69 & Philippines & AME (ESA)  \\
34 & Slovenia & NAPE  & 70 & South Korea  & AME (ESA)  \\
35 & Croatia  & NAPE  & 71 & Japan & AME (ESA)  \\
36 & Czech Republic & NAPE & & \\
\hline
\end{tabular}
\caption{List of countries and their assigned groups. Countries in each group are indexed according to the increasing longitude of their geographic centroids.}
\label{tab:country_groups}
\end{footnotesize}
\end{table}

\clearpage

\section*{CRediT authorship contribution statement}
Nahyeon Lee: Software, Formal analysis, Investigation, Writing – original draft.
Jongsoo Lim: Investigation, Supervision.
Hyeong-Chai Jeong: Investigation, Writing – original draft, Supervision.

\section*{Declaration of competing interest}
The authors declare that they have no known competing financial interests or personal relationships that could have appeared to influence the work reported in this paper.

\section*{Data availability}
 Data will be made available on request.

\section*{Acknowledgments}
 This work was supported by the National Research Foundation of Korea (NRF) grant,
 funded by the Ministry of Science and ICT (MSIT) (No. RS-2024-00359230)
 and the Ministry of Education (No. NRF-2022S1A5A2A03051182) of the
 Republic of Korea.
 
\section*{
Declaration of generative AI and AI-assisted technologies in the writing process
}

During the preparation of this work the authors used ChatGPT (OpenAI) in order to improve clarity, grammar, and style. After using this tool, the authors reviewed and edited the content as needed and take full responsibility for the content of the publication.


\begin{thebibliography}{10}
\bibitem{global_definition}
Global dataset represents the TV show that was top-ranked in the largest number
  of countries (among the 71 in our study) on each day.
\end{thebibliography}

\end{document}